\documentclass[seceq]{ptptex}

\usepackage{graphicx}

\newcommand{\la}{\langle}
\newcommand{\ra}{\rangle}
\newcommand{\lla}{\la}
\newcommand{\rra}{\ra}

\newcommand{\beq}{\begin{eqnarray}}
\newcommand{\eeq}{\end{eqnarray}}

\renewcommand{\theequation}{\thesection.\arabic{equation}}

\newcommand{\btem}{\bibitem}

\title{
Low Mass Scalar Mesons and Related Topics%
}

\author{
Teiji \textsc{Kunihiro}\footnote{ e-mail address:
kunihiro@yukawa.kyot-u.ac.jp}  
}

\inst{
Yukawa Institute for Theoretical Physics,
Kyoto University, Kyoto 606--8502, Japan
}

\abst{
We give a brief survey on the physical significance of the low-mass scalar
mesons in QCD, and also report on recent lattice studies on the sigma and 
kappa mesons. The importance to explore the in-medium properties of
the hadrons is mentioned.
}

\begin{document}
\maketitle

\section{Introduction; physical significance of low-lying scalar mesons}

A new era of hadron spectroscopy may have begun,
 which was  initiated by the announcement of the pentaquark baryon
\cite{nakano} and many other possible exotic hadrons with  mass larger than 2 GeV 
containing heavy quarks\cite{Quigg:2005tv}.
These experimental developments prompted the intensive theoretical
studies of QCD dynamics with new as well as old ideas on the structure
and dynamics of the hadrons, which include
chiral dynamics,
multi-quark states with diquark correlations or
molecular states and hybrids\cite{Quigg:2005tv}. 
It is to be noted that such a controversy on the structure of hadrons is
also the case  for the scalar mesons below 1 GeV; they are
 the $I=0$ and $J^{PC}=0^{++}$ meson, i.e.,
the $\sigma(400-600)$\cite{PDG} seen in the $\pi$-$\pi$ scattering 
and in decay products from heavy-quark systems\cite{KEK,Bugg}
and also the iso-doublet strange scalar meson, i.e.,
the $\kappa$  with mass $\sim 800$ MeV 
 \cite{Bugg}.

The significance of the $\sigma$ in hadron physics may be
summarized as follows\cite{kuni95}: 
(i) The $\sigma$ with a mass $400\sim 600$ MeV is 
responsible for the intermediate range attraction in
the nuclear force; without the $\sigma$ contribution, any nucleus
can not be bound, nor possible our existence.
(ii) The existence of the $\sigma$ resonance 
can account for the $\Delta I=1/2$ enhancement in the decay process
K$^0\rightarrow 2\pi$  
in comparison  with K$^{+}\rightarrow \pi^{+}\pi^{-}$\cite{mls}.
(iii) The empirical value of the $\pi$-N sigma term 
$\Sigma_{\pi N}= (m_u+m_d)/2 \cdot \la N\vert \bar{q} q\vert N\ra \sim$ 
40-60 MeV may be a reflection of  
 the properties of the $\sigma$ as a collective mode:
In fact, 
an analysis\cite{hk90,kuni95}
 with a  chiral model which describes the $\sigma$ meson as 
a collective mode gives an enhancement of the scalar charge of the nucleon as
$\la N\vert \bar{u}u+\bar{d}d\vert N\ra \sim 9$,  
 which is almost sufficient  for accounting the empirical value
of the sigma term  with $(m_u+m_d)/2\sim 5.5$ MeV, 
in contrast with the naive quark model which would give
$\la N\vert \bar{u} u+\bar{d}d\vert N\ra =3$. 
 See Ref.\citen{kuni95}
for more detailed discussions on the significance of the $\sigma$ meson in hadron 
physics.

The basic idea underlying the present report  is
 that  the low-energy hadron physics
may be regarded as  a study of the nature of QCD vacuum.
 In other words,
hadron physics is a condensed matter physics
of the QCD vacuum: In this point of view, 
  hadrons are elementary excitations on top of the non-perturbative
vacuum, although  QCD itself is  written solely in terms of
 quark and gluon fields.
Such a viewpoint  on the vacuum in quantum field theories  
was introduced by  Nambu\cite{nambu1960}.
Thus one sees that the iso-scalar and scalar meson can have an important aspect
as a Higgs boson of QCD\cite{kuni95}:
The chiral transition is  a phase transition of 
 QCD vacuum with 
 $\la \bar{q}q\ra$ being the order parameter, as clearly shown
by the lattice simulations\cite{karsch02}.
In fact, Nambu\cite{nambu1960}
showed that  
 an isoscalar-scalar meson with $J^{PC}=0^{++}$,
i.e., the $\sigma$ 
emerges as a quantum fluctuations of the chiral order parameter
$\lla :(\bar{q}q)^2:\rra$ with the mass $2\, M_q$  as another 
collective mode as the pion does, where $M_q$ is the dynamically
generated quark mass. 
This picture should remain valid for QCD and
the NJL model in fact works rather well for describing some aspects of the 
low energy hadron dynamics related to chiral symmetry and its
 dynamical breaking\cite{HK94}.

If a phase transition is of second order or {\em weak} 1st order,
there exists  ``soft'' modes which decreases it mass when the
system approach the critical point;
the soft modes are actually  fluctuations of the order parameter
of the phase transition.
 For chiral transition,
the relevant fluctuation is described by $\la(:\bar{q}q:)^2\ra$,
which has the same quantum numbers as 
the $\sigma$-meson does, i.e., $(I=0, J^{PC}=0^{++})$.
Accordingly the $\sigma$ meson can become
 the soft mode of chiral transition at $T\not=0$ and/or
$\rho_B\not=0$\cite{ptp85,HK94}:
 $m_{\sigma}\rightarrow 0$, $\Gamma_{\sigma}\rightarrow 0$
The lattice calculation\cite{karsch02} of the  generalized mass 
$m^{\rm gen}_{\sigma}=\chi_{\sigma}^{-1/2}$
defined in terms of the scalar correlation function
$\lla (\bar{q}q)^2\rra$ shows that the above picture is valid.
In this respect, it is to be noticed that the peak position of the
correlation function or the fluctuation $\lla (\bar{q}q)^2\rra$
is used to identify the critical temperature of the chiral transition $T_c^{\chi}$,
which is known to coincide with the critical temperature for the deconfinement
$T_c^{\rm dec}$ given from the peak position of the fluctuation of the
Polyakov loop $\lla L\rra$\cite{karsch02}.
  
\section{Low-lying scalar mesons in lattice QCD}

As mentioned in the previous section,
there are controversies on the nature
of the low-lying scalar mesons.
 In the non-relativistic constituent quark model,
 $J^{PC}=0^{++}$ is realized as a $^3$P$_0$ state, which 
implies that the mass of the $\sigma$ should be
in the 1.2-1.6 GeV region.
Therefore some mechanism is needed to down the mass.
The possible mechanisms so far proposed include:
(1)~ The color magnetic interaction 
between the di-quarks as advocated by Jaffe\cite{jaffe}; according to this
conjecture, the $\sigma $ and other low-mass scalar mesons are
tetra-quark states.
(2)~ The collectiveness of the scalar mode as the pseudoscalar mode; 
a superposition of $q\bar q$ states, which collectiveness is
due to chiral symmetry\cite{nambu1960}.
(3)~ These  scalar  mesons may be simply a resonance states of
the NG bosons as  the
unitarized chiral dynamics could account for the 
existence of them\cite{OOR}.

Facing these problems with the $\sigma$,
it would be interesting to explore the possible existence and the
nature of the low-lying scalar mesons in 
the first-principle calculation of QCD.
 Alford and Jaffe\cite{Alford}
examined  whether the diquark correlations  in the $\sigma$  channel
is significant and thereby tried to have a suggestion 
the $\sigma$ is a tetra-quark state.
They found in fact a large attraction for heavy quark systems.
But it should be warned that their calculations 
do not include the disconnected diagrams, i.e.,
closed quark loops, which  means that the state being calculated is not a genuine
flavor-singlet state.  
The SCALAR collaboration\cite{SCALAR} performed a full QCD calculation of the
$\sigma$ meson using the hybrid Monte Carlo method, which incorporated 
explicitly the disconnected diagrams by the $Z_2$-noise method with the
$8^3\times 16$ lattice: The Wilson fermion with three quark masses and 
plaquette gauge action
are employed; see \citen{SCALAR} for the details of the lattice setup.
The full QCD calculation includes the q-$\bar{\rm q}$ creation and annihilation
processes into the pure gluon states in the intermediate states so that 
possible tetra-quark, glue-ball states and so on can be taken into account,
in contrast to the quenched approximation.
The results are summarized as follows: The better signal of the $\sigma$ propagator
is obtained for smaller quark masses and the contribution from the
disconnected diagrams dominate the propagators over the connected
ones.  The simulation shows the existence of a clear
$\sigma$ resonance, especially for  smaller quark masses,
and the resulting $\sigma$ is almost degenerated with 
the $\rho$ meson for the smallest quark mass, 
although the  $\sigma $ mass becomes  much smaller than 
the $\rho$ mass when the naive chiral limit is taken.
Wada et al\cite{wada} have
 recently performed a lattice calculation of the $\kappa$ meson mass
in the quenched level with a large lattice to see whether the flavored scalar meson
to  which the disconnected diagrams do not contribute can have a small mass as
800 MeV as obtained in experiment. The result is negative as
anticipated, which may
mean that the scalar mesons including the $\sigma$ and the $\kappa$ 
should have exotic structures 
which can not be described by the simple constituent quark model.

\section{Summary and concluding remarks}

We have emphasized that 
the $\sigma$ meson and other low-lying scalar mesons are as
mysterious as the $\Theta^+$ and other candidates of the exotics with charm:
Naive quark model is in trouble for explaining
such a low-mass state in the $^3P_0$ state;
it may be a four-quark or  $\pi$-$\pi$
resonance state with no internal quark structure. 
The $\sigma$ might be also  a collective q-$\bar{\rm q}$
state to be identified as  the quantum fluctuation of the
 order parameter of the chiral transition as Nambu originally suggested.
The existence of such a collective mode in the scalar channel
can account for some  phenomena in hadron physics
which otherwise remain mysterious.
In short, 
the understanding of the nature or the even (non-)existence of the
low-lying scalar mesons, especially of the $\sigma$,
is important for a deep understanding of the QCD vacuum 
as well as the QCD/hadron dynamics.

A full QCD lattice simulation suggests the existence of a low-lying sigma, 
though its physics content, i.e., a tetraquark, a hybrid 
with the glue ball or the q-$\bar{\rm q}$ collective state, is obscure:
The disconnected diagram gives the dominant contribution to 
the $\sigma$ propagator. The simulation shows that 
$m_{\pi} <m_{\sigma}<m_{\rho}$ in the chiral limit.
A quenched Lattice calculation suggests that the $\kappa$
 can not be a normal q-$\bar{\rm q}$ state, either.

 To identify the nature of the $\sigma$ meson, 
exploring the possible change of the spectral function 
in the scalar channel in the hot and/or dense medium 
would be interesting, especially to examine
whether the $\sigma$ meson  can be
really identified with the quantum fluctuations of the chiral order
parameter\cite{HK94,ptp85,HKS,jido};
a peculiar enhancement of 
 the spectral function in the $\sigma$ channel 
in the lowering energy side  may be observed
along with a partial restoration of chiral symmetry in the medium\cite{jido}.

Recently, possible  $N_c$-dependence of the nature of the $\sigma$ meson 
has been noticed by some authors\cite{schafer};
T. Schaefer showed that 
at $N_c=3$ the low mass $\sigma $ exists which is described as a linear combination
 of q-$\bar {\rm q}$  and $({\rm q}\bar{\rm q})^2$.
However, for larger $N_c$, $m_{\sigma}$ goes up and the $\sigma$ becomes
 mainly composed of q-$\bar{\rm q}$.
The same problem is examined  by Pelaez
but somewhat different conclusions are deduced.
The fate of a hadron in the large $N_c$ limit might also tell whether
the hadron is an ordinary hadron as a Feshbach resonance or an
extraordinary hadron\cite{jaffe2}.

\section*{Acknowledgements}
This report include  the results obtained in the lattice calculation
done by the SCALAR collaboration.  I am grateful to all
 the members of the SCALAR collaboration for the collaboration.
This work was supported  by the Grant for Scientific Research (No.17540250) 
and by the Grant-in-Aid for the 21st Century COE 
``Center for Diversity and Universality in Physics"
of Kyoto University.




\begin{thebibliography}{99}

\bibitem{nakano}
  T.~Nakano {\it et al.}  [LEPS Collaboration],
  Phys.\ Rev.\ Lett.\  {\bf 91} (2003) 012002.

\bibitem{Quigg:2005tv}
For reviews, see, C.~Quigg,
  PoS {\bf HEP2005} (2006) 400;
E.~S.~Swanson,
  Phys.\ Rept.\  {\bf 429}, (2006) 243.

\bibitem{PDG} { Particle Data Group} Collaboration, S. Eidlman {\it et al}., Phys.\ Lett.\ {\bf B592} (2004) 1.
%
\bibitem{KEK} For example, see {\it Possible existence of the sigma-meson 
	and its implications to hadron physics}, KEK Proceedings 2000-4;
E791 Collaboration,
Phys.\ Rev.\ Lett.\ {\bf 86} (2001) 770. 
%
\bibitem{Bugg} 
D.~B.~Bugg, Phys.\ Lett.\ {\bf B572} (2003) 1;\,
E791 Collaboration, Phys.\ Rev.\  Lett.\ 
{\bf 89} (2002) 121801;\,
 {BES} Collaboration,
Phys.\ Lett.\ {\bf B633} (2006) 681.
%
\btem{kuni95} Teiji Kunihiro,  Prog. Theor. Physics.  Suppl. {\bf 120} (1995), 75.
%
\btem{mls} T. Morozumi, C. S. Lim and I. Sanda,
Phys. Rev. Lett. {\bf 65} (1990), 404.
%
\btem{hk90}
T. Kunihiro and T. Hatsuda, Phys. Lett.  {\bf B240} (1990), 209; \, 
T. Hatsuda and T. Kunihiro,
 Nucl. Phys. {\bf B387} (1992), 715.
%
\btem{nambu1960} Y. Nambu, Phys. Rev. Lett. {\bf 4}, 380 (1960);\,
Y.~Nambu and G.~Jona-Lasinio, Phys.\ Rev.\ {\bf 122} (1961) 345; 
{\bf 124} (1961) 246. 
%
\btem{karsch02} F. Karsch, Lect. Notes in Phys. {\bf 583} (2002), 209.
%
\btem{HK94}
T.~Hatsuda and T.~Kunihiro, Phys.\ Rep.\ {\bf 247} (1994) 221.
%
\btem{ptp85}T. Hatsuda and T. Kunihiro,
  Prog. Theor. Phys. {\bf 74} (1985), 765;\, 
 Phys. Rev. Lett. {\bf 55} (1985), 158; 
 Phys. Lett. {\bf B185} (1987), 304;\,
 the proceedings of IPN
Orsay Workshop on Chiral Fluctuations in Hadronic Matter,
September 26- 28, 2001, Paris, France, {\tt nucl-th/0112027}.
%
\bibitem{jaffe}
R.~L.~Jaffe, Phys.~Rev.{\bf D15}, 267, 281 (1977);\,
see also D.~Black et al
 Phys.~Rev.{\bf D59} (1999) 074026;\,
L.~Maiani et al
Phys.~Rev.~Lett. {\bf 93} (2004) 212002.
%
\btem{OOR}See, for instance, J. A. Oller, E. Oset and A. Ramos,
Prog. Part. Nucl. Phys. {\bf 45}, 157 (2000).
%
\bibitem{Alford} M.~Alford and R.~L.~Jaffe, 
Nucl.\ Phys.\ {\bf B578} (2000) 367;
see also K.~ F.~ Liu, these proceedings.
%
\bibitem{SCALAR} SCALAR Collaboration, 
Phys.\ Rev.\ {\bf D70} (2004) 034504.
%
\btem{wada} H. Wada et al, hep-lat/0702023.
%
\btem{HKS} T. Hatsuda, T. Kunihiro and H. Shimizu, Phys. Rev. Lett.
 {\bf 82} (1999), 2840;\,
 D. Jido et al, Phys. Rev.{\bf D63}
 (2001), 011901;\, 
 K. Yokokawa et al, Phys. Rev. {\bf C66}  (2002), 022201.
%
\btem{jido} D. Jido, these proceedings and the references cited therein.
%
\btem{schafer}T. Schafer, Phys. Rev. {\bf D68}, 114017 (2003);\,
J. Pelaez, Phys. Rev. Lett. {\bf 92}, 102001 (2004).
%
\btem{jaffe2}R.~L.~ Jaffe, hep-ph/0701038 and  these proceedings.
\end{thebibliography}
\end{document}